\documentclass[twocolumn,epjc3]{svjour3}  
\usepackage{latexsym}
\usepackage{graphics}
\usepackage{graphicx}
\usepackage{comment}
\usepackage{amsmath}
\usepackage{setspace}
\usepackage{lineno}
\usepackage{color}
\usepackage[numbers]{natbib}
\usepackage[breaklinks=true]{hyperref}
\hypersetup{
    colorlinks=true,
    linkcolor=blue,
    citecolor=blue,
    filecolor=magenta,      
    urlcolor=cyan,
    pdftitle={},
    }

\journalname{Eur. Phys. J. A}
\begin{document}

\title{AGATA DAQ-box: a unified data acquisition system for different experimental conditions
}


\author {
A.~Korichi\thanksref{ijclab,e1}, E.~Cl\'ement\thanksref{ganil}, N.~Dosme\thanksref{ijclab}, E.~Legay\thanksref{ijclab}, O.~St\'ezowski\thanksref{ip2i}, A.~Goasduff\thanksref{LNL}, Y.~Aubert \thanksref{ijclab,e2}, 
N.~Dosme\thanksref{ijclab}, J.~Dudouet\thanksref{ip2i}, S.~Elloumi\thanksref{ijclab}, P.~Gauron \thanksref{ijclab}, X.~Grave\thanksref{ijclab,e3}, M.~Gulmini\thanksref{LNL}, J.~Jacob\thanksref{ijclab}, V.~Lafage\thanksref{ijclab}, 
P.~Le Jeannic\thanksref{ijclab}, G.~Lalaire\thanksref{ganil}, J.~Ljungvall\thanksref{ijclab}, C.~Maugeais\thanksref{ganil},   
C. Michelagnoli\thanksref{ganil,ILL}, R.~Molini\thanksref{ijclab}, G.~Philippon\thanksref{ijclab}, S.~Pietri\thanksref{gsi}, D.~Ralet\thanksref{gsi, ganil,ijclab,e4}, M.~Roetta\thanksref{LNL},  F.~Saillant\thanksref{ganil},  M.~Taurigna-Quere\thanksref{ijclab}, N.~Toniolo\thanksref{LNL}
}
\thankstext{e1}{e-mail: amel.korichi@ijclab.in2p3.fr}
\thankstext{e2}{Deceased}
\thankstext{e3}{The author is on leave from IJCLab-CNRS}
\thankstext{e4}{Present address MIRION technologies France}

\institute{
IJCLab, IN2P3-CNRS, Universit\'e Paris-Saclay, bat 104-108, F-91405 Orsay Campus, France \label{ijclab}  \and
GANIL, CEA/DRF-CNRS/IN2P3, BP 55027, 14076 Caen cedex 5, France \label{ganil} \and
IP2I, IN2P3-CNRS, Universi\'e Lyon 1, France \label{ip2i} \and
LNL, INFN-Legnaro, Italy \label{LNL} \and
Institut Laue-Langevin, B.P. 156, F-38042 Grenoble cedex 9, France \label{ILL} \and 
GSI, Helmholtzzentrum für Schwerionenforschung GmbH, Darmstadt, Germany \label{gsi}
}
\date{Received: date / Accepted: date}
\maketitle

\abstract{
The AGATA tracking detector array represents a significant improvement over previous Compton suppressed arrays.
The construction of AGATA led to numerous technological breakthroughs in order to meet the requirements and the challenges of building a mobile detector across Europe. This paper focuses on the design and implementation of the data acquisition system responsible of the readout and control of the germanium detectors of AGATA. Our system is highly versatile, capable of instrumenting AGATA and seamlessly adapting it to various configurations with a wide range of ancillary detectors and/or spectrometers. It consists of three main components: an autonomous and independent infrastructure, a dedicated application core ensuring overall consistency, and a high--performance software package providing a fully integrated data flow management including the setting-up, the supervision and the slow control of the instrument. In this paper, we present a comprehensive analysis of the system's design and performance, particularly under high-counting rate conditions.
\PACS{ {07.85.Ne}{$\gamma$--ray tracking} \and
     {29.30.Kv}{$\gamma$--ray spectroscopy}  \and
     {29.40.Gx}{Data Flow, NARVAL, DCOD, Topology Manager}
     } 
}

\section{Introduction}

The most advanced implementations of the concept of gamma-ray tracking to date are the two arrays AGATA (Advanced GAmma Tracking Array) and GRETINA (Gamma Ray Energy Tracking In beam Nuclear Array)~\cite{AGATA_NIM,gretina,kor_tl19}. 
These arrays are built from large, segmented crystals of hyper-pure germanium and will be the first ones to use the concept of $\gamma$-ray energy tracking. 
The tracking concept is based on the ability to locate, within a few mm, each photon interaction point in the Ge detector and, consequently, to track the scattering sequence of an incident photon through the crystals. 
The method consists in the reconstruction of the full $\gamma$-ray energy by combining the appropriate interaction points\cite{kor_tl19}.

This approach naturally results in a significant gain in detection efficiency over escape-suppressed arrays because the Compton suppression shields (which limit the Ge solid angle) are removed and replaced by active Ge detectors. 
For the first time, a near 4$\pi$ sphere of Ge, with a good peak--to--total ratio, becomes possible. 
Moreover, the tracking technique provides identification of the first interaction point with good angular resolution and, therefore, allows for an improved Doppler correction. 
The expected performances are thus well beyond those of escape-suppressed spectrometers such as EUROBALL~\cite{simpson97} and Gammasphere~\cite{Lee90,janssens96}, enabling experiments probing low cross sections and/or measurements using high--velocity reaction products such as those possible with stable and radioactive beams at new facilities.
The resolving power of a $\gamma$-ray detector array ({\it i.e.}, its ability to isolate a given sequence of $\gamma$ rays in a complex spectrum) depends on four main properties~\cite{radford92}: efficiency, energy resolution, peak--to--total ratio $(P/T)$ (the ratio of photo-peak to the total efficiency~\cite{knoll79}), and granularity. 
The AGATA and GRETINA arrays, and the future $4\pi$ array GRETA~\cite{kor_tl19,AGATAandGRETA}, are being designed to maximize each of these properties. Their performance relies heavily on their high-rate capabilities in terms of front-end electronics (FEE) and data acquisition, enabling the investigation of low cross-sections using high beam intensities. To effectively handle the significant amount of raw data generated, a robust on-line processing is integrated into the data acquisition framework of AGATA. This on-line processing encompasses various tasks, including Pulse Shape Analysis (PSA), merging AGATA data with complementary instruments based on time-stamps, implementing the Tracking algorithm, and optimizing disk storage.  
The implementation of this comprehensive approach leads to a significant reduction in overall data volume, thereby ensuring efficient handling and processing of the data. The primary objective is to prevent data flow transport from becoming a limiting factor in AGATA's on-line performance. Furthermore, this approach enhances the performance of the data processing algorithm through the utilization of large-scale task distribution and parallel processing.
The data acquisition system must establish a cohesive and self-consistent environment for both hardware and software topology, encompassing AGATA and its complementary instrument. 
This includes data flow coupling, run-control, state-machine management, metadata handling, and parameter management, essential for on-line processing, near--online checks, and off-line analysis across the entire array, up to a 4$\pi$ system consisting in 180 AGATA crystals.
These challenges assume even greater significance given AGATA's nature as a mobile detector, designed to operate across various European facilities and configurations. 

Maintaining a high resolving power, exceptional performance, versatility, reliability, and the ability to collect reusable data, even in demanding high-activity environments, is of utmost importance for AGATA.
To achieve these objectives, an innovative data acquisition system called DAQ-box was developed with three key fundamental components which will described in this paper. Furthermore, to meet the demands of high--performance instruments like AGATA, substantial computing, networking, and storage capabilities are required, which will be extensively discussed in this contribution.

While not providing all technical details, we aim to provide important elements for a broad audience.
We will briefly discuss the 3 fundamentals of the DAQ-box, show an overview of the whole system in Sec.~\ref{s:overview}, followed by a description of the computing infrastructure in Sec.~\ref{s:infrastructure}, and the Topology Manager's relevant functionalities for AGATA in Sec.~\ref{s:topologyM}. The performance capability of the whole system will be presented in Sec.~\ref{s:performance}, while the associated software package features and developments are detailed in  Ref.~\cite{ref-software} in this volume. Some repetitions will be made to guide the reader in understanding this complex topic.
\section{Fundamentals of the AGATA DAQ-box}
\label{s:fundamentals}
In order to attain the objective of high rate capability (50 kHz/crystal) for a 4$\pi$ array as well as accommodate the extensive computational power and large data transfer bandwidth, the AGATA data acquisition system (DAQ) is built upon a distributed architecture  based on NARVAL\footnote{NARVAL stands for Nouvelle Aquisition temps-Reel Version Avec Linux meaning New Aquisition in Real time Version Alongside Linux}\cite{grave05}. To address the requirements and challenges, the AGATA DAQ-system, incorporates three key software features. 
Firstly, NARVAL/DCOD enables high data flow, ensuring efficient handling of the generated data. Secondly, a specific Slow/Run Control system is implemented to fulfill the requirements of the host laboratory and coupled ancillary detectors. Lastly, a Topology Manager ensures the versatility of the detector while maintaining a high level of integrity.
Additionally, it employs a high-performance infrastructure consisting of a sophisticated network operating at high bandwidth, numerous HTC (High-Throughput Computing) workstations, and extensive disk storage devices. NARVAL integrates on-line PSA, $\gamma$-ray Tracking, and data-analysis tools, enabling real-time experiment optimization.
More details on these tools can be found in Ref.~\cite{ref-software,ref-psa,ref-tracking} in this volume.

Furthermore, AGATA's construction phases are anticipated to undergo multiple upgrades in the forthcoming years, which can be seamlessly incorporated into the DAQ-box system. The design of the DAQ-box enables smooth adaptation to emerging technologies, including GPUs (Graphics Processing Units), and facilitates the integration of innovative algorithms, such as those utilizing Machine Learning for the PSA and Tracking. Successfully meeting these challenges is only possible with DCOD (Distributed Caen Orsay DAQ)\cite{dcod_wiki,grave18} approach.

\section{General overview of the Data Flow: from AGATA crystals to storage}
\label{s:overview}
This section offers an overview and  detailed description of the algorithms seamlessly integrated within the NARVAL data flow, along with their associated components at both the Local and Global levels. We also highlight the supplementary software features surrounding DCOD. Furthermore, we offer a concise explanation of the transition from NARVAL to DCOD, which serves as a foundation for a comprehensive introduction to the Topology Manager.
\begin{figure*}
\centering 
    \includegraphics[width=0.8\textwidth]{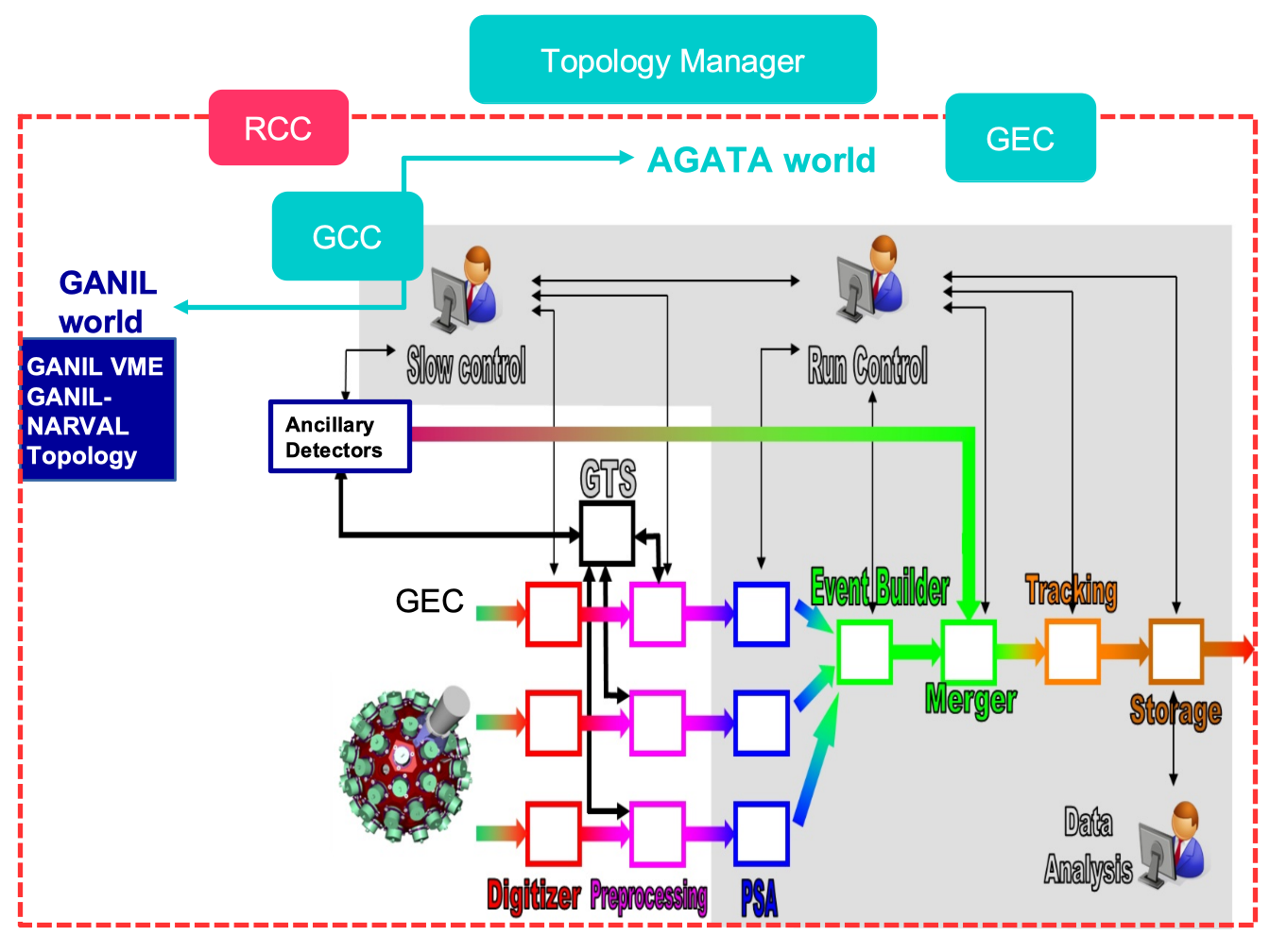}
\caption{\label{fig:layout2} 
The data acquisition layout of AGATA, as implemented at GANIL, highlighting the DAQ-box (depicted in the grey area) and the distinctive supplementary components exclusive to the host laboratory (see text for more details).
}
\end{figure*}
In AGATA, the data flow is designed in a way that treats each crystal as an independent entity. Each crystal consists of a central contact and 36 segments, which are handled as separate entities by the front-end electronics. The NARVAL data flow software incorporates a "Local Level Processing" approach that maintains this separation throughout the entire process until the $event- builder$ stage. At the event builder stage, individual crystal data are merged based on their respective time-stamps making the transition to the Global Level Processing. 


\subsection{Data Flow system main architecture}

The front-end electronics generate an output of 200 Bytes per channel by sampling 100 points at 10 ns intervals from the digitizer, with each point requiring 16 bits (2Bytes)\footnote{During the Local Level Processing stage, each event occupies 7.6 kBytes. This can be calculated by multiplying the number of channels (38) by the output size per channel (200 Bytes).}.This results in a data bandwidth of 380 MBytes/s/cystal when operating at a rate of 50 kHz per crystal, which includes the 36 segments and high/low gain central contact signals. This throughput is efficiently managed by the first layer of NARVAL data flow software. 
 
Communication between the FEE and the PSA computers employs Gigabit Ethernet interfaces, while the data flow integrates algorithms necessary for processing information from the interaction points.

NARVAL data flow system consists of three types of coordinated processes designed to handle data from the detectors:  Producers (handling incoming data), Intermediaries (filters, mergers) and Consumers (data storage into files, histograms). Together, these coordinated processes form the core of the NARVAL data flow system, ensuring efficient handling and processing of data from the detectors.
Fig.~\ref{fig:layout2} shows the AGATA data flow structure (depicted in the grey area) as implemented in phase 1 and the distinctive supplementary components exclusive to the host laboratory. These additional components significantly enhance the specific setup at GANIL and will be discussed in the upcoming text. This layout form the foundation for the phase 2 construction, ensuring further advancements and improvements.

The data flow consists of six distinct classes:
\begin{itemize}
\item{FEE (Digitizers and Pre-processing modules) catcher}
\item{Pulse Shape Analysis (PSA)}
\item{Event Builder (EB) of the AGTA crystals}
\item{Merger with Ancillaries}
\item{Tracking algorithm}
\item{Data Storage}
\end{itemize}

After the PSA, the data volume is reduced, and all resulting information's, including positions and energies of the interaction points, are merged together by the EB taking into account the relevant physics correlations provided by the pre--processing electronics. These correlations are based on a 10-ns resolution timestamp. If additional detectors are coupled to the AGATA array, the corresponding data flow is assembled in the same way by the Merger. After this procedure, Tracking is performed to reconstruct the $\gamma$-ray trajectories. This task is then followed by the Data Storage on a large local disk array before being sent to Grid Tier1 computing centres based at INFN-CNAF in Bologna (Italy) and at CC-IN2P3 in Lyon (France). It is also possible to write data to disk at each step of the data flow\footnote{Note that when possible, e.g  for low multiplicity $\gamma$-rays --with AGATA triggered by an ancillary detector resulting  in a typical rate of $\sim$ 100 Hz/crystal, the traces from the FEE can be stored hence contributing to a significant amount of data of $\sim$3TB/day. However, when dealing with high spin physics that generates high multiplicity $\gamma$-rays, this approach becomes impractical or even unattainable.}.

The event-builder and merger algorithms are written in Ada~\cite{ada} while the PSA and Tracking algorithms are written in C++ language. The originality of the DAQ system relies on the development of the ADF library, which has been designed in order to facilitate the separation between data transport and algorithm developments. ADF provides the interface between the data flow 
and the algorithms (PSA and Tracking) and allows one to encode/decode the data at various stages of the flow. ADF can be either used with NARVAL or as a stand-alone library for data replay \footnote{ The replay process allows the users to optimize the post-PSA or tracking or to use new PSA algorithms that are being developed in the collaboration when the traces are stored.} and it is part of the software package.
NARVAL incorporates a surveillance feature to monitor the DAQ system, enabling sampling of data fragments from each actor. This 'spy' mechanism grants control, monitoring, and online analysis capabilities and is accessible to any client connected to the DAQ services network.

\subsection{From NARVAL to DCOD}
In 2017, NARVAL was upgraded to DCOD \cite{dcod_wiki,grave18} and was implemented into AGATA at that time. Even if this upgrade results with almost no impact on the users during the experiments, the fundamental changes in the buffer managements improved the rate capabilities of the data (when the PSA is not performed) by a factor of two.  Moreover, this upgrade was necessary for the phase 2 of the project when new technologies, such as GPU's, will be integrated in the DAQ-box as soon as the new electronics
(see Ref. \cite{ref-electronics} in this volume, for more details) is complete.

\subsection{Local Level Processing and Global Level Processing}
\begin{figure}[htb]
\begin{center}  
    \includegraphics[width=0.5\textwidth]{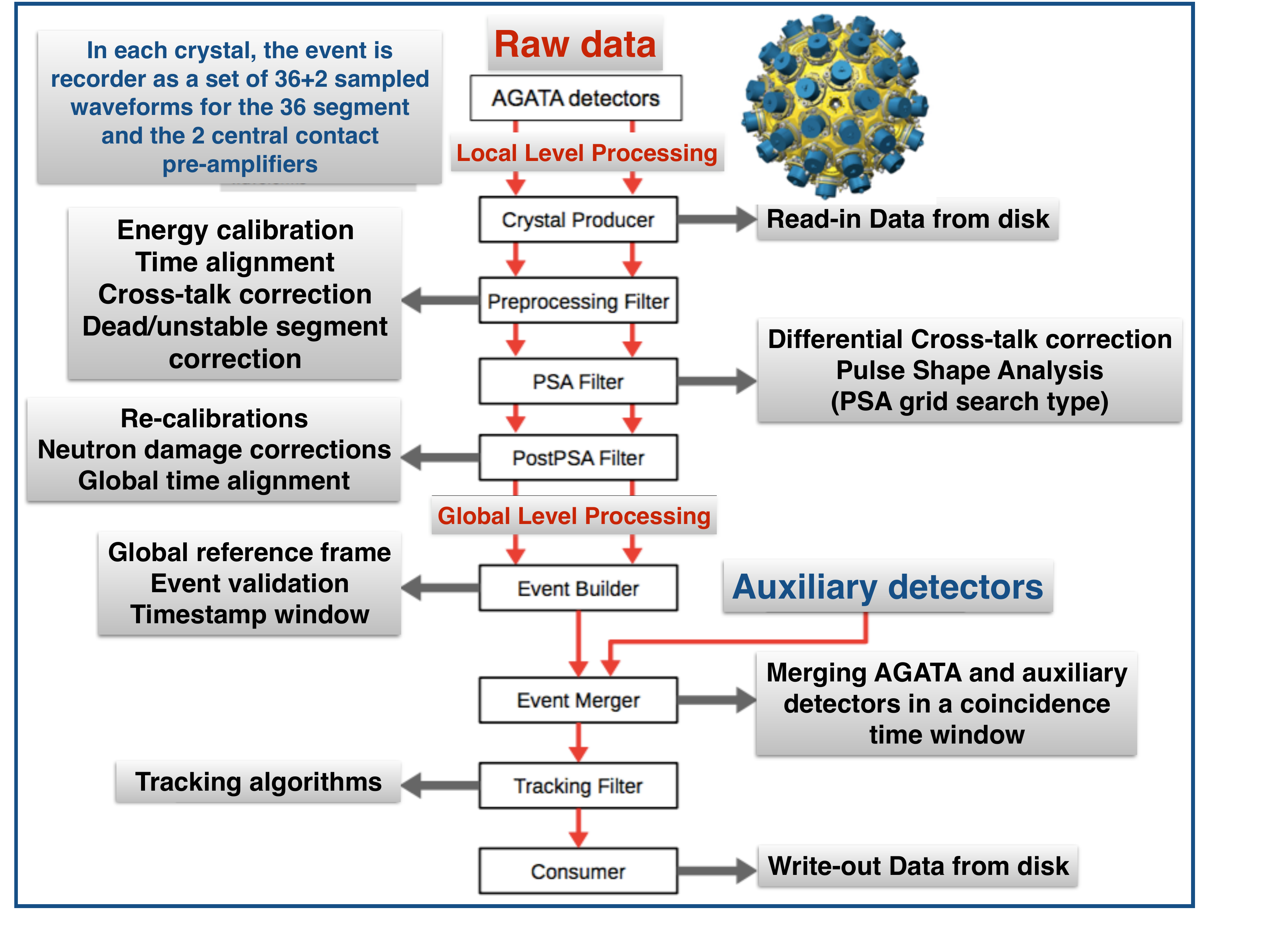}
\end{center}
\caption{\label{fig:narval_chain}
Full NARVAL chains and theirs corresponding actors, filters and consumers as employed throughout the complete data processing workflow.
These components collectively illustrate the intricate calibration and processing procedures implemented within the system.
}
\end{figure}

The complete DCOD chains and their corresponding actors are illustrated in Fig.~\ref{fig:narval_chain}, depicting the sequential steps involved in the processing/treatment of raw data from the top level down to the Tracking and Storage stage.

It maps the procedures from the pre-processing to the storage as shown in  the bottom half of Fig.\ref{fig:layout2}. At the local level processing, one can see the incoming data from the FEE being treated by the Crystal Producer actor which is followed by the Pre-processing Filter in which the first step in the treatment of raw data
for each AGATA crystal are performed. In other words, it shows all the complex data treatments and calibration procedures for which several filters have been designed at the local or global level and includes the consumers that can be defined at any step in order to write the data and histograms to disk. For more information, please refer to Ref.\cite{ljungvall20}.
When accounting for all the parameters associated with the FEE and filters incorporated within the DAQ-box for these data treatments and calibration, each AGATA crystal entails the management of $\sim$1700 parameters. To scale this up to the final 4$\pi$ system, it becomes necessary to handle approximately 300,000 parameters.

\subsection{Supplementary Software Components surrounding DCOD}
In addition to the previously mentioned general descriptions, the collaboration has implemented various new features to enhance the functionality of the system. These include the establishment of an independent network separate from the host laboratory and the development of hardware and software bridges. These enhancements allow the system to operate autonomously and seamlessly integrate with ancillary particle detector and recoil spectrometers. 
However, it is important to acknowledge that the AGATA DAQ-box may require modification and/or upgrade when deployed at a specific facility. 
For instance, during the GSI physics campaign, the coupling between the local MBS (Multi Branch system) DAQ and AGATA data flow was necessary; but required significant efforts from both AGATA and GSI teams~\cite{ralet15}. The architecture undergoes continuous modifications when AGATA is relocated between different sites. A similar adaptation was undertaken during the GANIL campaign, leading to the development of the depicted layout showcased in Figure.~\ref{fig:layout2}.

To ensure the successful integration of AGATA within the host laboratory, a set of vital software tools were developed that revolve around the DCOD data flow. Among the key software components are: 
\begin{itemize}
\item{GEC, Global Electronics Control for the whole AGATA system dispatching the state machin of the AGATA subsystems}
\item{GCC, Global Control Core acting as a software bridge between the RCC and DCOD which enables the preparation/start and stop of the data acquisition}
\item{RCC, Run Control Core acting as "chef d'orchestre" orchestrating the readout and data processing for both AGATA and complementary detector actors, effectively distributing the run ID and other meta data of the collected data}
\item{Topology Manager which ensures the coherency of the whole system and provides consistent information to the configuration files that are needed by different entities of the whole system (from the electronics to the DAQ configuration)}
\end{itemize}
The RCC plays a crucial role in distributing and collecting the data acquisition flow, current status of actors, user instructions, and run metadata in a coherent manner. It serves as the central hub for transmitting this information to the GCC for the AGATA sub-system, as well as to the data flow of complementary detectors from LNL, GSI, and GANIL. It is important to note that the RCC is not a singular entity but rather a versatile component that adapts to the host environment of AGATA.

In the phase 1 of AGATA, the RCC was provided by the host laboratory and, with a simple API (Application Programming Interface)\cite{API}, it is translated by the GCC server to the AGATA world. With this feature, the common {\bf Start, Stop, Kill} call is performed simultaneously for both AGATA and ancillary data flow. While reading/accessing the AGATA topology XML (Extensible Markup Language)\cite{xml} files and the host laboratory data flow description file, the RCC gets the whole topology of the experiment as generated by Topology Manager described in Sec.~\ref{s:topologyM}. It also assigns a unique run ID and distributes the corresponding timestamp for the data identification. Beside this, it can provide a detailed description of the electronics chain for each run number. 
It is worth noting that the RCC--manages are based on different data flow interfaces at the host laboratories: At GANIL it is based on NARVAL~\cite{clement_design}, MBS at GSI~\cite{ralet15} and XDAQ at LNL~\cite{gadea_design}. This complex task together with the complete RCC monitoring, is only possible due to the flexibility and the fundamental conception of NARVAL and DCOD data flow system.

\section{AGATA DAQ-box Infrastructure: towards a fully autonomous and portable system}
\label{s:infrastructure}
\begin{figure}[htb]
\begin{center} 
    \includegraphics[width=0.5\textwidth]{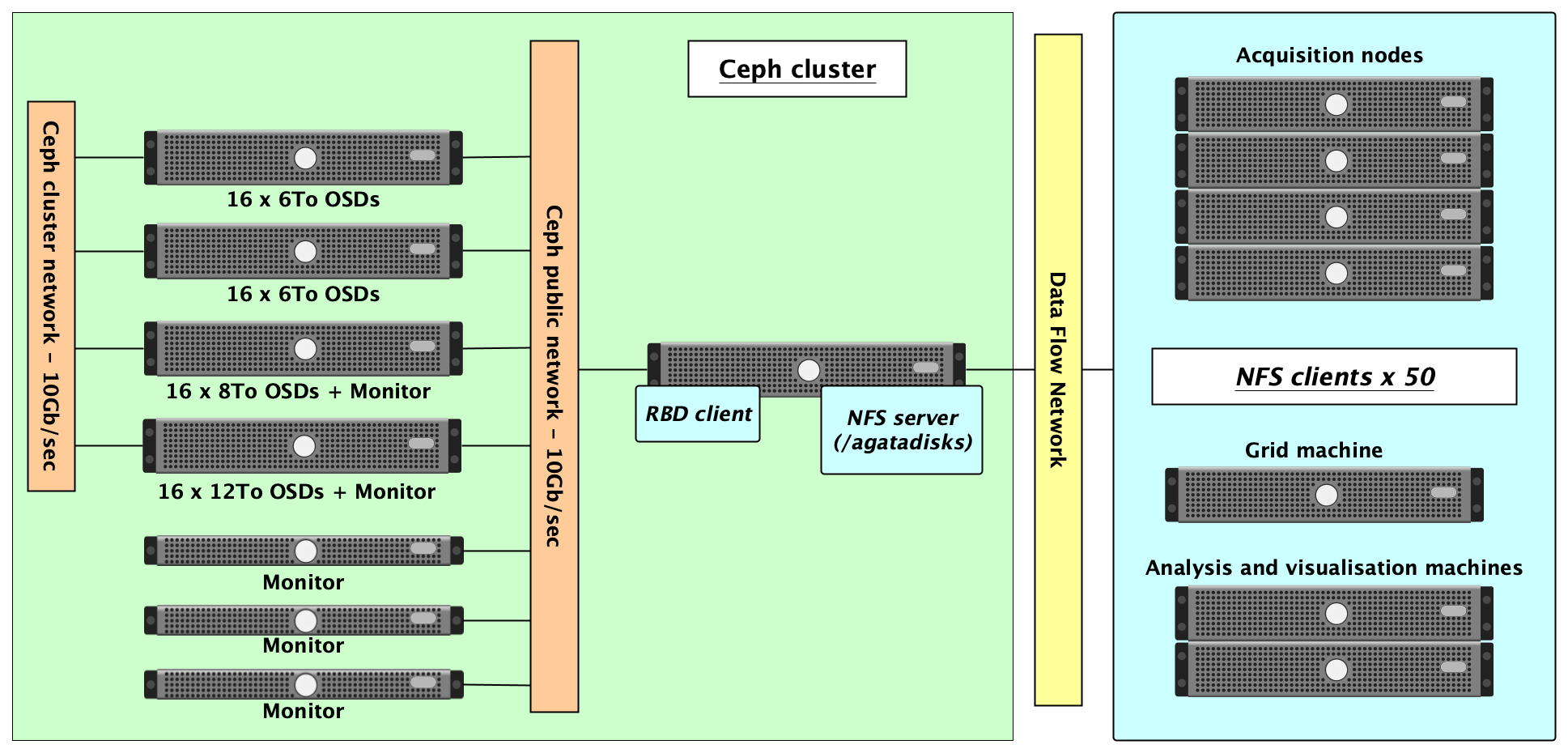}
\end{center}
\caption{\label{fig:ceph} 
Ceph storage architecture at LNL-Legnaro and its links to the data acquisition nodes, data analysis and Grid machine.
}
\end{figure}

The AGATA DAQ-box infrastructure is a complex object and its configuration evolves while populating the array with more crystals/detectors. In the early phase~2, the pre-processing electronics is hosted in the computing nodes. As shown in Fig.~\ref{fig:layout2}, this implies that each digitizer is attached to a 2U-server for the PSA and results in a number of nodes equal to the number of crystals in the array.
\begin{figure}[htb]
\begin{center}  

    \includegraphics[width=0.5\textwidth]{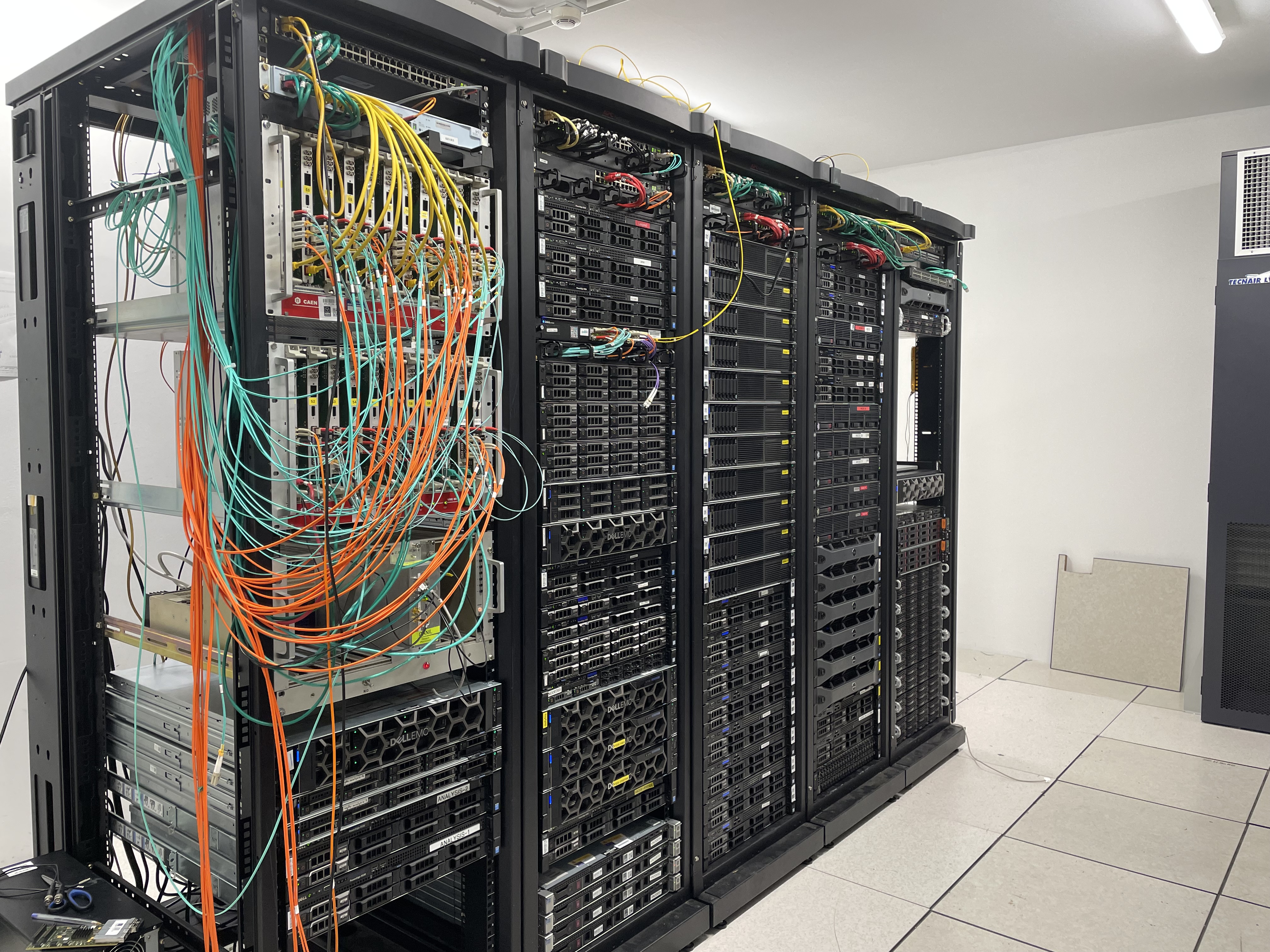}
\end{center}
\caption{\label{fig:pic} 
This photograph shows the portable AGATA-DAQ box currently installed at LNL (Italy). The leftmost rack holds the AGATA clock distribution hardware and the analysis servers. The second rack contains all the services and disk servers. The remaining racks are dedicated to the computing nodes for the crystals.
}
\end{figure}
The  pipeline  component  pattern is adapted to multiple--core CPUs by incorporating additional  threads  running  whatever  elements  are  most  critical.  In  the  case  of  PSA,  this  is  the  central  computing  element~\cite{grave_GTAG2}. Beside this, four nodes are devoted to the data analysis for the near--online control of the data integrity and quality, and 8 additional nodes are used for the storage. 
The AGATA storage infrastructure utilizes the Ceph cluster technology\cite{ceph-cern} . It employs a distributed storage architecture consisting of four disk nodes, housing a total of 64 disk units, providing a massive storage capacity of 466 TB (see Fig.~\ref{fig:ceph}). This storage system is commonly referred to as the AGATA Tier0.

To ensure data integrity and prevent loss, the online disk server employs a triple redundancy mechanism, which allows for $\sim$ 130 TB of available space for data taking. As previously stated in the introduction, AGATA data is consistently transferred\footnote{The current management of incoming and outgoing bandwidth involves the implementation of four distinct queues, each with a maximum limit of 300 MB/s. These queues operate in parallel and are dynamically managed to prioritize low latency, particularly during the execution of concurrent large data transfers.} in a "duplicate mode," ensuring that two separate Grid Tier1 storage computing centers receive copies for enhanced safety measures. This process is carried out following nearly every experiment to enable users to conveniently access and process the data on the Grid. Concurrently, it effectively frees up the capacity of the current online Ceph storage, optimizing resource allocation for improved system performance.
In terms of the associated networks, beside the small LAN (Local Area Network) for the clock distribution system configuration, four specific networks are dedicated to the data, the services and the Ceph storage. 
The AGATA network is isolated from the host laboratory network, ensuring enhanced security measures. It can be accessed through a secure OPNSense gateway that hosts a variety of services, all which can be configured based on the specific experimental requirements. 
Unlike the GANIL phase, the AGATA gateway has undergone expansion for the ongoing LNL campaign. It now includes the management of the the coupled ancillary/complementary instrumentation. The latest additions are now regarded as an integral part of the AGATA sub-network. To facilitate remote control and monitoring of the entire array, a dedicated unique and secure VPN connection is employed. This consolidated infrastructure ensures unified and protected remote access, enabling efficient management and oversight of the AGATA system.
It is important to note that the current architecture will undergo significant modifications with the installation of the phase 2 electronics~\cite{ref-electronics} and these changes will have a profound impact on the system. 
Indeed, the phase 2 data flow will rely on Ethernet 10 Gb links between the FEE and the DAQ-box eliminating the need for direct point-to-point electronic board connections. This transition offers several benefits and advantages, including the ability to migrate from a 2U server/crystal configuration to a high-performance computing farm, while still using DCOD. Consequently, the DAQ system's footprint will be significantly reduced. Additionally, using Ethernet links between the FEE and the DAQ-box allows for longer distances to be covered without significant signal degradation, making it an attractive option for applications with remote digitizer locations.
Furthermore, the 10-Gb FEE outputs can be concentrated and transmitted through a 100-Gb high-speed link, resulting in an improved data transmission speed and efficiency.

Moreover, the upcoming upgrade to a more flexible high--performance computing farm presents exciting opportunities for architectural advancements. Through the implementation of  multi-parallel and multi-threading PSA processing techniques, leveraging CPU, CPU+GPU, or GPU technologies, AGATA's event rate capabilities are expected to experience substantial enhancement. The enhanced computational power and flexibility provided by this upgrade hold the potential for  significant improvements in AGATA's overall performance and scientific outcomes.
\section{Topology Manager: towards a fully consistent DAQ-box}
\label{s:topologyM}

\begin{figure}[htb]
\begin{center}  
\centering
    \includegraphics[width=0.5 \textwidth]{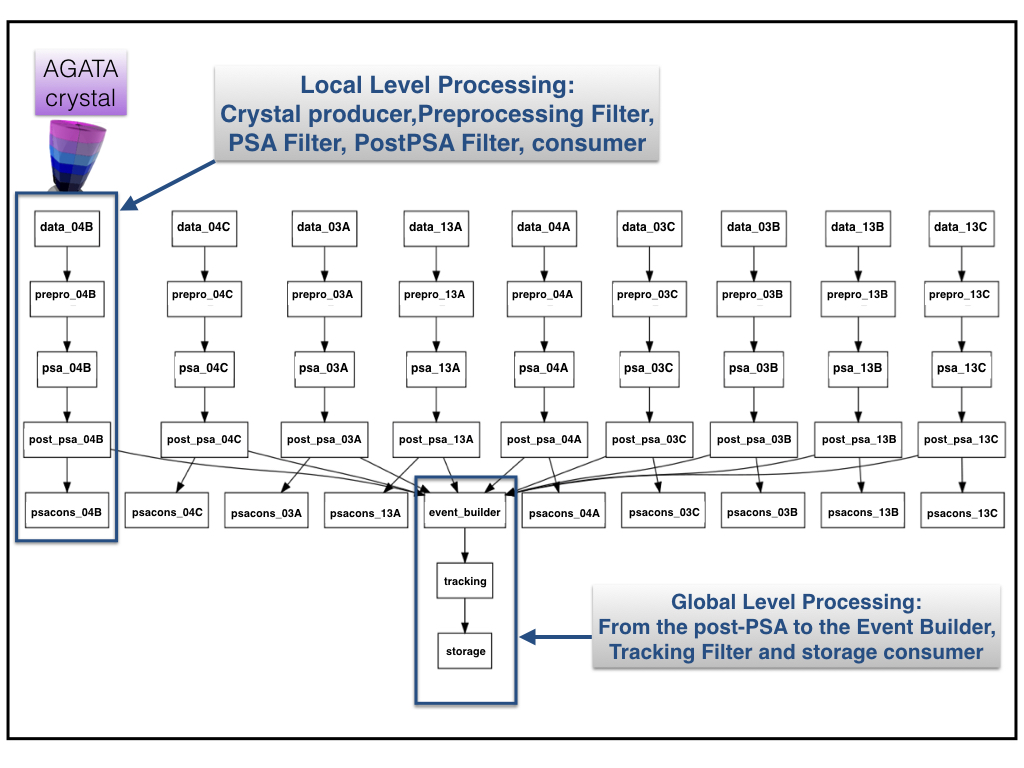}
\end{center}
\caption{\label{fig:fm} 
Topology Manager: example of topology generation for 9 AGATA crystals with the global level processing (see text for more details). 
}
\end{figure}
During the data taking, NARVAL/DCOD actors can run on any node of the DAQ-box, and the configuration is established using the Topology Manager (TM) at the outset. 
The web interface provides users with a convenient means to effortlessly activate or deactivate individual crystals and their respective acquisition chains. It also enables the generation of DAQ and electronics configuration files without requiring extensive knowledge of the system architecture.
Data buffers are exchanged between actors, and their size is determined by the TM based on the acquisition rate and resulting data bandwidth. This approach ensures efficient data transfer and maximizes system performance. Additionally, the TM performs a thorough consistency check of the user--defined configuration when generating the set of files, thereby minimizing the likelihood of array configuration errors and enhancing system performance.
 

\subsection{FEE configuration}

\begin{figure}[htb]
\begin{center}  
    \includegraphics[width=0.49\textwidth]{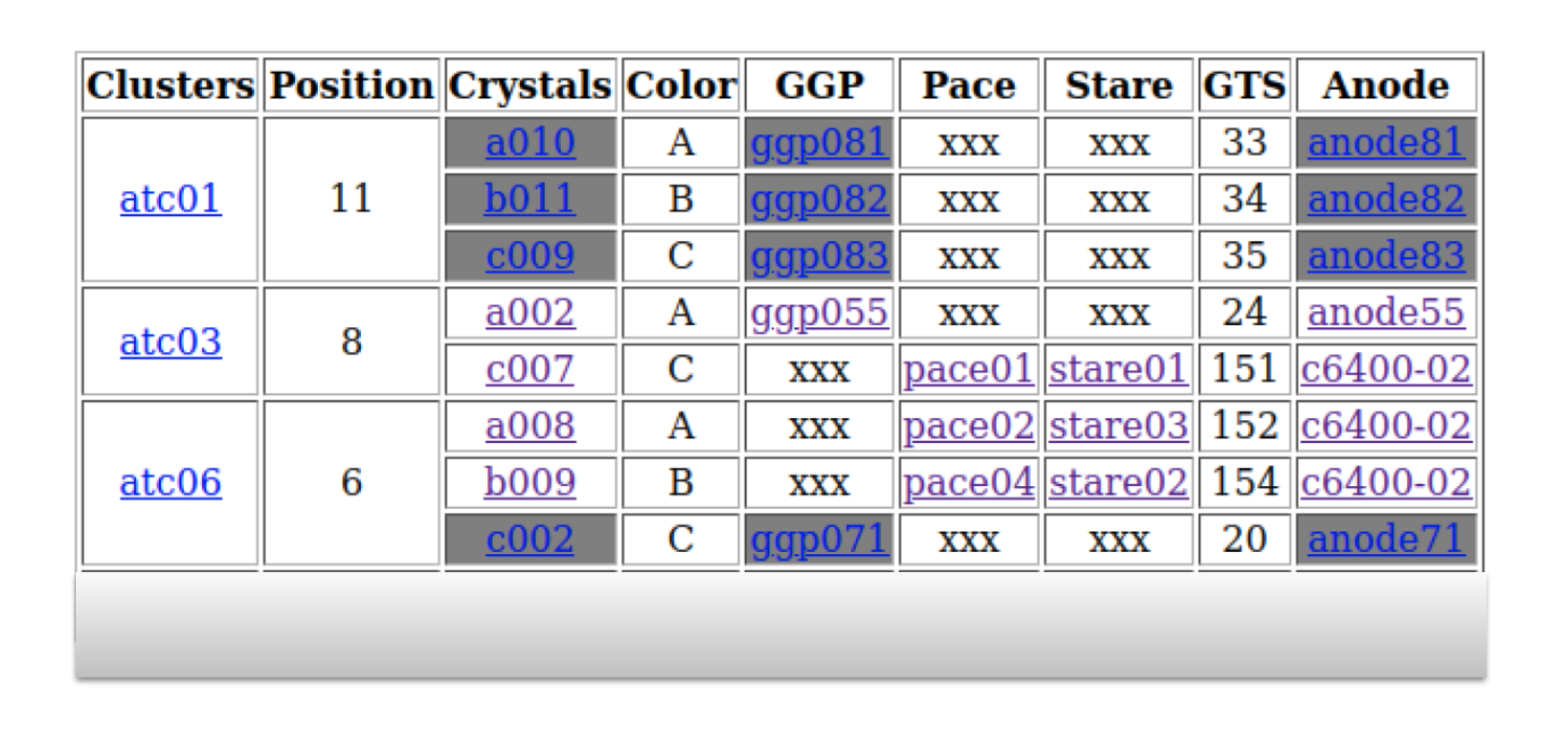}
\end{center}
\caption{\label{fig:FEE_TM} 
Example of the FEE Topology Manager combining the phase 1 (GGP) and the upcoming phase 2 (PACE/STARE) electronics~\cite{ref-electronics}.
}
\end{figure}

The AGATA clock and trigger system, also known as the GTS~\cite{gts}, needs to have a specific list of hardware and channels included in the topology for proper function. A sample topology is presented in Fig.~\ref{fig:FEE_TM}, where each AGATA Triple Cluster detector (atc01 for example) with its three crystals (a010, b011 and c009) is allocated a unique hardware material (ggp081, ggp082, ggp083). The GTS channels, along with the electronics boards GGPs (Global Gigabit Processor)and computers (Anode for AGATA node), are exclusively assigned for each crystal.

For the phase 2 of AGATA, the GGP electronic board, the GTS clock together with the trigger distribution will change (Please refer to Ref.~\cite{ref-electronics} for more information). The Topology Manager will consistently handle the pre-processing board called PACE~\cite{ref-electronics}, the readout board STARE, and the clock-trigger distribution SMART for all sub-services such as the GEC, RCC, and NARVAL/DCOD. 

\subsection{Generation of NARVAL/DCOD topologies}

The full acquisition chain is described in an XML file which includes the complete description of each individual box (actor) presented in the layout of Fig.~\ref{fig:fm}. For data transfer between actors, NARVAL/DCOD employs two modules, the PMH (Posix Memory Handler) to manage buffers and the CTL (Common Transport layer) to manage buffer's transfers between computers. The XML file includes:
\begin{itemize}
    \item the depth and dimension of input/output buffers;
    \item the connection to the other actors of the chain;
    \item the libraries to be loaded;
    \item paths to the configuration files;
    \item verbosity of each actor.
\end{itemize}
As mentioned earlier, the acquisition process/chain of a single crystal involves three essential actors: the producer, the pre-processing filter and the PSA filter. Additionally, data monitoring tools must be included, resulting in a total of six actors per crystal. For a complete AGATA 2$\pi$-configuration, 540 actors are required at the local level, highlighting the importance of an automated tool for generating all the necessary DAQ topology files. To simplify buffer size management, the TM offers the users a pre-defined configurations based on the acquisition rate. 
 
\subsection{Run Control}
The AGATA NARVAL/DCOD-box features a local run control called the Chef d'Orchestre (CO), as described in Ref.~\cite{grave05}. The CO centralizes commands, such as {\bf Configure, Start, Stop and Kill}, from the Host Laboratory Run Control (HLRC) and distributes them to all actors in the DAQ topology, significantly simplifying the coupling of the DAQ-box with existing local infrastructure. The HLRC does not need to interact with the NARVAL/DCOD XML topology files. The CO can also monitor actor status, buffer status, and data bandwidth through SOAP (Simple Object Access Protocol) communication. In the current LNL implementation of AGATA, the HLRC periodically requests all monitoring information, which is then sent to a time-series database, enabling online buffer status monitoring and better dimension adjustments. Communication with the FEE, particularly enabling/disabling data transmission from the FEE to the DAQ, is handled by the HLRC. To simplify HLRC tasks, environment variables are created by the TM listing all active crystals, corresponding nodes, and electronics.

\section{Performance of the DAQ-box}
\label{s:performance}
The performance of the AGATA system including the DAQ-box, is continuously evaluated based on the specific requirements of the host laboratory according to its specific conceptual design \cite{clement_design, gadea_design,pardo_design,dobon_design}. 

A high--performing tracking array should produce spectra with high energy resolution, a good P/T ratio and with high efficiency. It must also be able to determine the first interaction point for $\gamma$ rays within a few mm to minimize the effect of Doppler broadening for fast moving nuclei. Additionally, it must perform well under conditions of high counting rates as well as high multiplicity events for high--spin physics. To meet these requirements, the DAQ-box was designed to sustain high rates, such as 50 kHz/crystal. A detailed study of the relative efficiency (for dead-time evaluation) and resolution of an AGATA detector as a function of the count rate has been reported in Ref.~\cite{recchia}. 

\begin{figure}[htb]
\begin{center}  
    \includegraphics[width=0.49\textwidth]{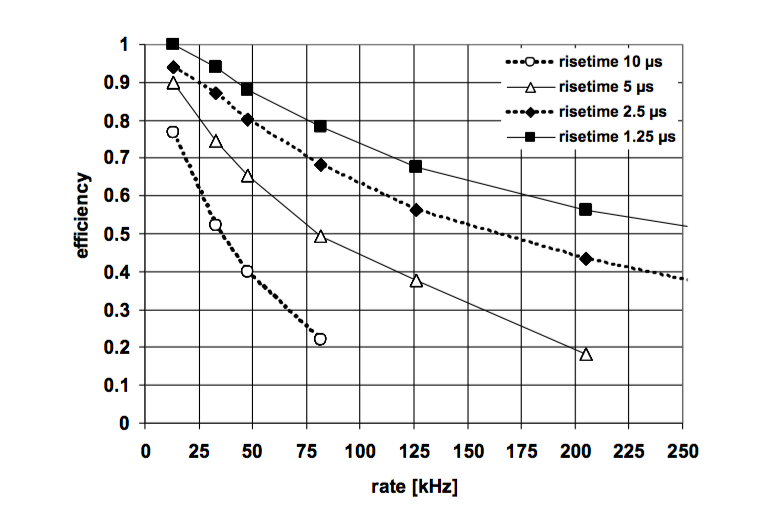}
\end{center}
\caption{\label{fig:risetime} 
Relative efficiency as a function of the counting rate for different MWD rise-times. Taken from Ref~\cite{recchia}.
}
\end{figure}

Fig.~\ref{fig:risetime} demonstrates that when the trapezoidal filter's integration time~\cite{trapeze} (referred to as "rise time") is appropriately adjusted at 50 kHz, the relative efficiency consistently stays above $\sim$80$\%$. Additionally, the measured resolution at 1.3 MeV is around 3 keV, which is considered highly satisfactory given the prevailing conditions~\cite{recchia}.
 
 Many other measurements for evaluating the absolute response function of the array, have also been performed, which are not discussed in this paper, but can be found in Refs.~\cite{natasa16,caterina16,korichi17b,ljungvall20,assie21}. 

\begin{figure}[htb]
\begin{center}  
    \includegraphics[width=0.49\textwidth]{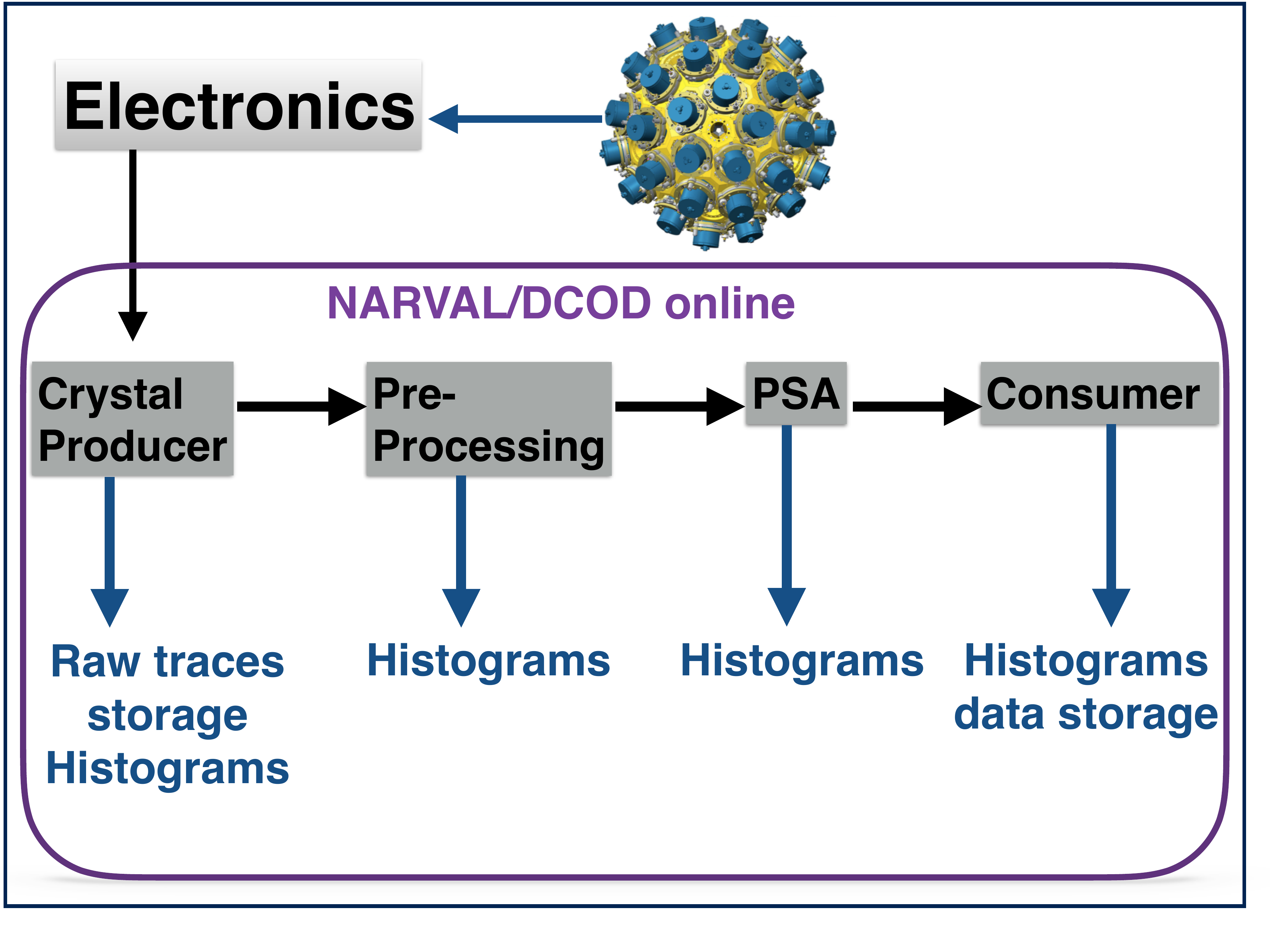}
\end{center}
\caption{\label{fig:dcod_online} 
A simplified depiction of the NARVAL/DCOD online topology, highlighting areas where rate bottlenecks may arise.
}
\end{figure}

A simplified topology of NARVAL/DCOD online is presented in  Fig.~\ref{fig:dcod_online} to illustrate where bottlenecks in rates might occur, with different levels of consideration including the crystal producer, pre-processing, PSA, and histogram consumers.
The DAQ--box features and level limitations have been extensively evaluated in various investigations~\cite{ak_GTAG2}. 
It's important to note that the acquisition rates are highly dependent on experimental conditions, including the number of crystals in the array. As a result, the performance of the data flow is regularly checked whenever the number of channels is increased or when new generations of pre-processing boards, such as ATCA\footnote{ATCA stands for AGATA Telecommunications Computing Architecture board, which refers to the phase 1 electronic boards that are no longer in use.}, GGP, and PACE/STARE in the near future, are added, or when major releases of C++ actors are updated.

\begin{figure}[htb]
\begin{center}  
    \includegraphics[width=0.49\textwidth]{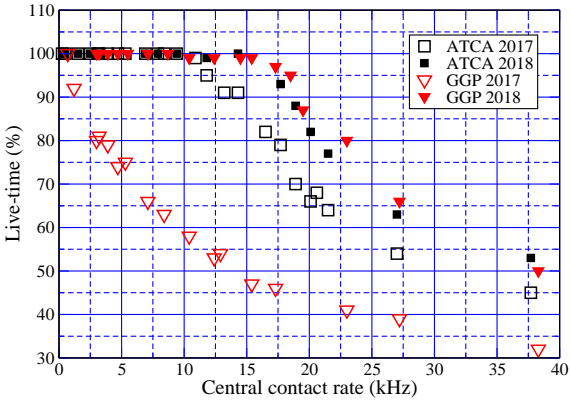}
\end{center}

\caption{\label{fig:lib17-18} 
Crystal producer efficiency as a function of rates for the phase 0 (ATCA) and phase 1 (GGP) electronic channels. NARVAL/DCOD readout libraries have been modified to enable the utilization of massive threads. The figure clearly illustrates the improvements and highlights the delicate process of PCIe readout in DCOD. 
The measurements were conducted on identical detectors, comparing the performance of the library before (2017) and after (2018) the modifications were made.
}
\end{figure}

To illustrate the importance of these evaluations, Fig.~\ref{fig:lib17-18} presents the system's live--time in terms of buffer event rates, as managed by the data flow from the PCIe memory block of the FEE to the DCOD consumer. This data was collected with DCOD in 2018 using high-rate radioactive sources positioned in front of the corresponding AGATA crystals. The distance between AGATA and the source was varied to gradually increase the rates per crystal. For these measurements, two generations of pre-processing boards (ATCA and GGP) and two Crystal Producers (as shown in Fig.~\ref{fig:lib17-18} ) were compared. It should be noted that the PSA rate was known to be limited to around 4 kHz and was therefore disabled for these data sets.

Fig.~\ref{fig:lib17-18} highlights the rate loss of the ATCA and GPP crystal producers (using the  2017 library) above 11 kHz and 1 kHz respectively. However, a major new release of the Crystal Producer was introduced in 2018 {\it by D. Bazzacco}~\cite{dino} resulting in a significant improvement of the GGP and ATCA readout capabilities. This now makes the performance for both pre-processing boards in terms of rate equivalent thanks to the exploitation of C++ multi-threading libraries, which can be absorbed without any difficulty by DCOD. 
The current limit is measured to be $\sim$17 kHz/crystal in a "loss-free mode event rate" for the two electronic chains. However, high rates cannot be handled without setting specific trigger conditions in order to avoid "dead--time" loss.

The AGATA data can be triggered by any ancillary device compatible with the GTS system, thanks to the hardware trigger processor. 
When large acceptance spectrometer such as VAMOS~\cite{vamos} or PRISMA~\cite{prisma} trigger AGATA, the validation rate per crystal is typically kept below a few hundreds Hz per crystal, even though the requested rates may exceed 50 kHz/crystal. This scenario is similar to that observed when using radioactive ion-beams with low to average beam intensities. However, regardless of the trigger conditions, caution must be exercised while adjusting the buffer dimensions, as poorly sized buffers may cause timeout issues and result in data loss.

\begin{figure}[htb]
\begin{center}  
    \includegraphics[width=0.5\textwidth]{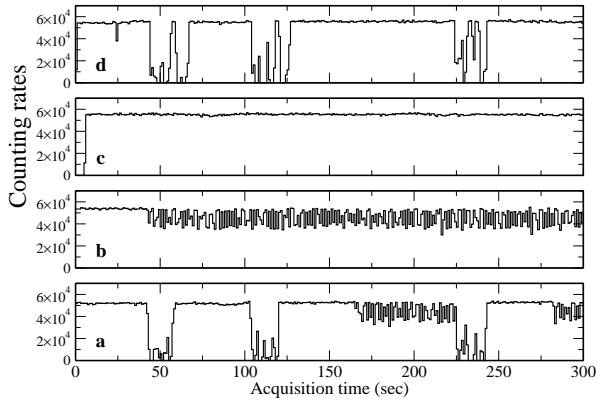}
\end{center}
\caption{\label{fig:rate_Er} 
Counting rates for high multiplicity $\gamma$--ray events in different conditions of the DAQ as a function of acquisition time. Panel a) corresponds to data taking for which the buffer size was not optimized while running at $\sim$ 20 kHz/crystal and collecting all histograms. Panel b) depict the same conditions as a) but without histogram collection. Panel c) demonstrates data acquisition with properly configured buffer size and without histogram collection. Panel d) mirrors c) but with all histograms enabled. 
}
\end{figure}

The impact of buffer size  on AGATA's data flow is demonstrated in Fig.~\ref{fig:rate_Er}.
These measurements were performed for $^{158}$Er for high-multiplicity events conditions with fusion--evaporation reactions. 
The nucleus was produced in $^{122}$Sn($^{40}$Ar,4n)$^{158}$Er reaction at 170 MeV with beams from the GANIL facility. 
The decay $\gamma$ rays were measured with AGATA (comprised of 24 crystals at the time). The experiment aimed to assess the system's overall response under stress. The same reaction has been used for the GRETINA tracking array response's function evaluation, with 28 crystals. 
Panel c) in Fig.~\ref{fig:rate_Er} depicts a smooth and stable data rate with no loss, obtained when the system was run without overload. Conversely, in panel d), the buffer size configuration for the producer, PSA and post-PSA was identical to that of panel c) during histograms collection : despite the similar buffer size, the system encountered an overload and resulted in different outcomes.
However, when writing the traces for high multiplicity data, the overload resulted in a significant loss of over 70$\%$. 

Additionally, when less selective ancillaries, such as the neutron detector array NEDA\cite{neda_agata}, were used,  the reduction factor between the requested and validated events was generally less than one order of magnitude compared with VAMOS or PRISMA--triggered data. 

As mentioned previously, the primary limitation in the AGATA data flow performance arises from the computing time of the PSA. By optimising buffer dimensions and keeping the system's "dead--time" below 10\%, an acquisition rate of approximately $\sim$4 kHz/crystal can be achieved without compromising the spatial resolution of AGATA. However, in a standalone mode and in particular when dealing with high multiplicity/high spin physics, the situation is different. NARVAL/DCOD can handle up to 50 kHz/crystal, but limitations arise from the disk access and the PSA. 

\begin{figure}[htb]
\begin{center}  
\includegraphics[width=0.5\textwidth]{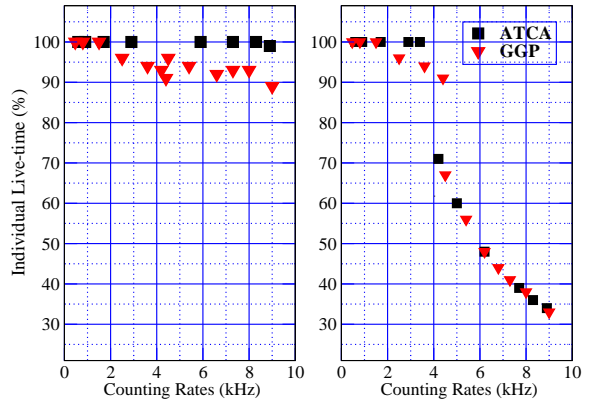}
\end{center}
\caption{\label{fig:psa_max} 
Individual live--times as measured for ATCA and GGP channels as a function of rates when the PSA is performed (right panel), compared with that of without PSA (left).
The performance of the system noticeably diminishes beyond 4 kHz, resulting in a significant loss of data, as evident from the decrease in the overall live-time of the system. Please note that this value should be considered as an average and is subject to potential variations based on the used hardware, as evident in  the comparison between GGP and ATCA.
}
\end{figure}

The effect can be observed in Fig.~\ref{fig:psa_max}, which demonstrates how the dead--time varies with counting rates when the PSA is activated. To measure this, radioactive sources were utilized under the same conditions as depicted in Fig.~\ref{fig:lib17-18}. In the left panel, data was collected for approximately 1 minute without activating the PSA actor, and no data loss was observed. Conversely, the right panel shows data collected with the PSA activated, resulting in data loss as the rates were increased. 
Currently, the PSA rate limit stands at around 4 kHz, using threading techniques with 5 threads, where each thread handles 300 events per crystal per node.
It is noteworthy that more recent HTC technologies have already surpassed this limit by a few additional kHz.



The phase 2 AGATA project has been defined and is exploring various avenues to process the PSA at high rates. 
According to Ref.~\cite{grave_GTAG2}, the new  Ethernet--based electronic boards, allow for the distribution of the CPU load over a high-performance computing farm, resulting in a remarkable achievement of over 10kHz/crystal PSA rate. Nevertheless, it is believed that some limitations may require significant optimization or modification of the existing algorithm implementation. One potential approach is to balance the PSA processing load by employing distinct PSA algorithms tailored to different level of event complexities. Additionally, memory access and cache may be a bottleneck for the PSA performance and are currently under evaluation ~\cite{vincent,romeo}.
In the future, with the introduction of new electronics, GPU optimization might be used to further improve the PSA processing within the Data Flow-system. 
Ref.~\cite{calore} investigated the utilization of GPUs and naturally, this approach needs to be tailored to meet the requirements of AGATA.
\section{Conclusion}
The AGATA collaboration has made significant efforts over the past decade to construct and operate the phase 1 of the project which offers the best $\gamma$--ray resolving power using tracking arrays. 
The challenging construction of the phase 2 is currently under way to complete a 4$\pi$ $\gamma$-ray tracking array.  
The data flow structure of AGATA will continue to use the NARVAL/DCOD architecture, which allows for memory access and network transmission to be managed by the memory POSIX handler and common transport layer.
This architecture offers the required flexibility, modularity, and robustness for the full AGATA array. 
With faster processors and today's algorithms, the processing of PSA at a rate exceeding 10 kHz per crystal becomes achievable. This can be accomplished by employing one or multiple anodes per crystal resulting in significant improvements in the overall performance of AGATA
These advancements enable AGATA to successfully detect the rarest events at upcoming heavy ion facilities currently under construction.
Load balancing and new technologies (multi-parallel processing, multi-threading, accelerated GPU calculation, IA and ML) also enable the use of different PSA algorithms depending on event complexity, further improving performance. The new electronic readout will be Ethernet-based, allowing for the CPU to be distributed over high-performance computer farms. In addition to infrastructure and performance benefits, a development of software trigger will represent a major improvement of the data flow for AGATA. 
These features are currently being developed or researched as part of the AGATA project definition framework.

\begin{acknowledgement}
We extend our heartfelt appreciation to Dino Bazzacco for his outstanding contribution to AGATA, particularly in the development of the AGATA DAQ-box. His invaluable guidance during the system upgrade, performance measurements, and data acquisition played a vital role in the success of AGATA operations.
We also acknowledge the valuable feedback and contributions from the users during the physics campaigns, which helped us make necessary improvements and achieve a robust DAQ-box.
Furthermore, we express our gratitude to the AGATA collaboration for their scientific support of this manuscript.
This work was made possible by the support of the French National Center of Research, CNRS, France.


\end{acknowledgement}
\bibliographystyle{apsrev4-1}
\bibliography{AGATA-DAQ_performance}

\appendix








%




\end{document}